\documentclass[reqno]{amsart}
\usepackage{amsmath,amsfonts,amssymb}
\newcommand{\be}{\begin{equation}}
\newcommand{\ee}{\end{equation}}

\def\f{\frac}
\def\p{\partial}
\begin{document}
\title{Generalized heat-transport equations: Parabolic and hyperbolic models}
\author{Patrizia Rogolino     \and
      Robert Kov\'acs   \and
			Peter V\'an \and
	  	Vito  Antonio Cimmelli
}



\address{
$^1$ Department of Mathematics and Computer Sciences, Physical Sciences and Earth Sciences,
		University of Messina, Viale F. Stagno d'Alcontres, 31, 98166, Messina, Italy    and 
$^2$Department of Energy Engineering, Faculty of Mechanical Engineering, 
				Budapest University of Technology and Economics, Budapest, Hungary and 
$^3$Department of Mathematics, Computer Science and Economics, University of Basilicata, Viale dell'Ateneo Lucano, 10, 85100, Potenza, Italy}

\date{Received: date / Accepted: date}

\maketitle

\begin{abstract}
We derive two different generalized heat-transport  equations:  The most general one, of the first order in time and second order in space, encompasses  some well known heat equations and describes the hyperbolic regime in the absence of nonlocal effects.
Another, less general, of the second order in time and fourth order in space, is able to describe hyperbolic heat conduction also in the presence of nonlocal effects.

We investigate the thermodynamic compatibility of both models by applying  some generalizations of the classical Liu and Coleman-Noll procedures.  In both cases, constitutive equations for the entropy and for the entropy flux are obtained.

For the second model, we consider a heat-transport equation which includes nonlocal terms and study the resulting set of balance laws, proving that  the corresponding thermal perturbations propagate with finite speed.

\end{abstract}
\section{Introduction}\label{Intro}
Heat-transport is currently enlarging its domain of applicability and discovering new phenomenologies in situations where the classical Fourier's theory is no longer applicable~\cite{Leb,SELCIMJOU}. Indeed, several new aspects arise as consequences of the relation between the heat carriers' mean free path $l$, and a relevant characteristic size of the system $L$,  represented by the Knudsen number $\operatorname{Kn}=l/L$. The Fourier's law  is valid in the limit of very small Knudsen number, i.e., when $l/L\ll1$. 
On the other hand, when generalized heat-transport equations are used, it turns out that the analysis of their consistency with the second law of thermodynamics requires a generalized framework, where the entropy and the entropy flux are not known a priori. In fact, even in simple situations, entropy and entropy flux are found to be more general than their local-equilibrium forms~\cite{CimJouRugVan}. 

Nowadays, there is a current interest for mesoscopic modelization, based on generalized heat-transport equations~\cite{Tzo1,TZOU,KovVan1} and weakly nonlocal thermodynamics~\cite{LebJouCasMus,CimSelJou1,CimSelJou2,CimSelJou3,VanFul,KovVan},  which is simpler than the much more complex and detailed microscopic approach. However, mesoscopic modelization applies different schemes and procedures, which derive by different thermodynamic theories~\cite{Cim3,CimJouRugVan} such as, Extended Irreversible Thermodynamics (EIT)~\cite{LebJouCasMus}, Rational Thermodynamics (RT) with internal variables~\cite{CimSelJou1}, Rational Extended Thermodynamics (RET)~\cite{DreStr}, Non-equilibrium Thermodynamics with internal variables ~\cite{VanFul,BerVan17b}. Thus, the development of a general method, which is independent of the particular thermodynamic theory, would be desirable.

The first attempt toward an universal procedure to obtain generalized heat-transport equations has been made in~\cite{VanFul}, and then in~\cite{KovVan}, where a general method, using internal variables and  a linear relationship between internal variables and entropy flux~\cite{Ver,Nyi},  has been exploited  to reproduce several models which are well known in literature.
The main assumptions in~\cite{VanFul} are the following:

\begin{itemize}
\item the deviation from the equilibrium state is conveniently characterized by a vectorial internal
variable; 
\item the deviation from the classical form of the entropy current is conveniently characterized
by an arbitrary tensorial function, called current multiplier.
\end{itemize}

In~\cite{KovVan}, instead, the previous assumptions have been generalized as follows: 

\begin{itemize}
\item the deviation from the equilibrium state is conveniently characterized by the heat flux and a second-order tensorial internal variable; 
\item the deviation from the classical form of the entropy current is conveniently characterized
by two arbitrary tensorial functions, called current multipliers.
\end{itemize}

The first approach is capable to reproduce both linear and nonlinear Maxwell-Cattaneo-Vernotte (MCV)~\cite{Cat1}  and Guyer-Krumhansl (GK)~\cite{GuyKru1,GuyKru2} equations, as well as Jeffreys type (JT)~\cite{Cim3}  and Green-Naghdi type (GNT)~\cite{GreNag} heat-transport equations.

Beside the mentioned models, the second approach allows to obtain also Cahn-Hilliard type~\cite{KovVan} transport equations.

In the present paper we go a step further and prove that, starting from Grad's 13-moments system~\cite{Gra},  
it is possible to obtain two different generalized heat-transport equations. The most general one, which encompasses  some well known heat equations, is capable to describe hyperbolic heat conduction only in the absence of nonlocal effects; 
another one, instead, obtained under additional special conditions, is able to describe hyperbolic heat conduction also in the presence of nonlocal effects.

In the first case, the basic unknown fields are the specific (per unitary volume) internal energy, the heat flux and the flux of heat flux. We suppose that the spatial derivatives of the unknown fields, included the dissipative fluxes, may enter the state space. As result, we get a very general system of equations. Its thermodynamic compatibility is investigated by applying the classical Liu procedure~\cite{Liu}, together with an Onsager method~\cite{Ons1,Ons2} for the solution of the reduced entropy inequality ~\cite{VanFul,KovVan,Van1,CimVan}.

In the second case, we consider some differential consequences of the previous system, by taking the space and time derivatives of its equations. By combining these derivatives with the original system, we obtain a higher-order evolution equation for the heat flux, which is of the second order with respect to time. We prove that this equation can be hyperbolic, leading so to finite speeds of propagation of thermal perturbations, even if the constitutive equation of the flux of heat flux is nonlocal, namely it depends on the gradients of the unknown fields.

A further problem which is worth to be investigated is the form of the constitutive equations of the entropy and of the entropy flux corresponding to both the situations presented above. To achieve that task we exploit second law of thermodynamics. We prove that:
\begin{itemize}
\item in the first case, the entropy is local and depends on the dissipative fluxes too, while the entropy flux is nonlocal;
\item in the second case, both the entropy and the entropy flux are nonlocal.

\end{itemize}

The paper has the following layout.

In Sec.~\ref{Equation1}, starting from the Grad's 13-moments system~\cite{Gra},  we exploit second law of thermodynamics to obtain a 13-moments system which is similar to that obtained in RET~\cite{DreStr,MR}. Then, we show that this system, derived in a very general way, encompasses several important heat-transport equations.

In Sec.~\ref{Equation}, we consider some differential consequences of Grad's 13-moments system, and derive a generalized heat equation of the second order in time and fourth order in space. The thermodynamic compatibility of the model is investigated by exploiting the entropy principle~\cite{CimJouRugVan}. To this end, we apply a generalized Coleman-Noll procedure~\cite{ColNol,CimSelTri1}, which is based on the substitution in the entropy inequality of the gradient extensions of the basic balance laws.  In this way, suitable constitutive equations for the entropy and for the entropy flux are obtained as consequences of second law of thermodynamics.

In Sec.~\ref{Wave}, we consider the system formed by the local balance of energy and a special form of the equation derived in Sec.~\ref{Equation}. We show that such a system is hyperbolic, i.e., it allows  the propagation of thermal perturbations with finite speed. 

In Sec.~\ref{Disc}, a comparison of the results obtained in Sects.~\ref{Equation1} and~\ref{Equation} is carried out. Possible future developments of the theory are discussed as well.

{\section{Generalized heat-transport equation I: Parabolic case}\label{Equation1}

Let's consider a one-dimensional rigid heat conductor, and let the coordinate $x$ denote the position of its points. The following system of balance laws is supposed to hold
\begin{eqnarray}
&&e_{,t}+ q_{,x}=0 \label{s1a}\\
&& q_{,t}+{\Phi}_{q,x}= r_q \label{s1b}\\ 
&&{\Phi}_{q,t}+\Psi_{,x}= r_p \label{s1c}
\end{eqnarray}
where the symbols $\{\cdot\}_{,x}$ and $\{\cdot\}_{,t}$ denote the partial derivative of the generic quantity  $\{\cdot\}$ with respect to $x$ and $t$, respectively. 

The system above represents the one-dimensional version of the 13-moments system of EIT~\cite{SELCIMJOU,CimJouRugVan,JCL}, which, in turn,  is directly amenable to  Grad's 13-moments system~\cite{Gra}, with $e$ as internal energy per unitary volume, $q$ as the heat flux, ${\Phi}_{q}$ as the flux of heat flux, $\Psi$ as the flux of ${\Phi}_{q}$, $r_q$ as the production of the heat flux $q$, and $r_p$ as the production of the flux of heat flux ${\Phi}_{q}$. The closure of~\eqref{s1a}-~\eqref{s1c} is achieved by assigning suitable constitutive equations for the flux $\Psi$ and for the source terms $r_q$ and $r_p$.

The present model is developed within the framework of weakly nonlocal thermodynamics~\cite{Van1}, according to which the spatial gradients of the unknown variables are allowed to be included in the state space. 

In this section we consider a first order weakly nonlocal state space and obtain a closure with  the help of Liu procedure~\cite{Liu},  using the above set of balances as constraints to entropy inequality. In the next section, instead,  we arrive to a hyperbolic set of equations with the help of a generalized Coleman-Noll procedure~\cite{ColNol,CimSelTri1}.

Therefore, let's take the state space 
\begin{equation}\label{2}
Z=\{e,  e_{,x}, q,  q_{,x}, \Phi_{q},  \Phi_{q,x} \}
\end{equation}
and the constitutive functions  $\Psi, r_q, r_p$, the entropy density $s$ and  the entropy flux $J$ defined on it. Introducing the Lagrange-Farkas multipliers $\lambda, \alpha, \beta$ for the balance equations \eqref{s1a}-\eqref{s1c} ~\cite{Van1}, we calculate the entropy inequality with the constraints \eqref{s1a}-\eqref{s1c} as follows~\cite{Liu}

\begin{gather}
s_{,t} + J_{,x} - \lambda(e_{,t}+ q_{,x}) -\alpha( q_{,t}+{\Phi}_{q,x}- r_q) -\beta({\Phi}_{q,t}+\Psi_{,x}- r_p) =
\nonumber\\
\f{\p s}{\p e} e_{,t} + \f{\p s}{\p e_{,x}} e_{,xt} + \f{\p s}{\p q} q_{,t} + \f{\p s}{\p q_{,x}} q_{,xt} + 
\f{\p s}{\p  \Phi_{q}} \Phi_{q,t} + \f{\p s}{\p \Phi_{q,x}}  \Phi_{q,xt} + \nonumber\\
\f{\p J}{\p e} e_{,x} + \f{\p J}{\p e_{,x}} e_{,xx} + \f{\p J}{\p q} q_{,x} + \f{\p J}{\p q_{,x}}q_{,xx} + 
\f{\p J}{\p  \Phi_{q}} \Phi_{q,x} + \f{\p J}{\p \Phi_{q,x}}  \Phi_{q,xx}  -\nonumber\\
- \lambda(e_{,t}+ q_{,x}) -\alpha( q_{,t}+{\Phi}_{q,x}- r_q) -\nonumber\\
\beta\left({\Phi}_{q,t}+
\f{\p \Psi}{\p e} e_{,x} + \f{\p \Psi}{\p e_{,x}} e_{,xx} + \f{\p \Psi}{\p q} q_{,x} + \f{\p \Psi}{\p q_{,x}}q_{,xx} + 
\f{\p \Psi}{\p  \Phi_{q}} \Phi_{q,x} + \f{\p \Psi}{\p \Phi_{q,x}} \dot \Phi_{q,xx} 
- r_p\right)\geq 0 \label{epLiu}
\end{gather}

Collecting the terms that are proportional to the time derivatives of the state variables, the solution of the corresponding Liu equations results in a local entropy density, that is $s = s(e,q,\Phi_q)$. Moreover, the Lagrange-Farkas multipliers are given by the partial derivatives of $s$ with respect to the basic fields, namely,  $\lambda =\f{\p s}{\p e} $,  $\alpha =\f{\p s}{\p q} $ and $\beta =\f{\p s}{\p \Phi_q} $. Finally, the coefficients of  the second order space derivatives of the basic fields, i.e. $e_{,xx}$,   $q_{,xx}$ and  $\Phi_{q,xx}$, must vanish as well, and this yields
\begin{eqnarray}\label{130a}
&&  \f{\p J}{\p e_{,x}} -  \f{\p s}{\p \Phi_{q}}  \f{\p \Psi}{\p e_{,x}} =0 \\ 
&&  \f{\p J}{\p q_{,x}} -  \f{\p s}{\p \Phi_{q}}  \f{\p \Psi}{\p q_{,x}} =0\\\label{130b}
&&  \f{\p J}{\p \Phi_{q,x}} -  \f{\p s}{\p \Phi_{q}}  \f{\p \Psi}{\p \Phi_{q,x}} =0 \label{130c}
\end{eqnarray}

The general solution of the partial differential equations \eqref{130a}-\eqref{130c} can be written as
\begin{equation}
J = J_0 +  \f{\p s}{\p \Phi_{q}} \Psi \label{efLiu}
\end{equation}
where $J_0$ is local, too, that is $J_0 = J_0(e,q,\Phi_q)$

Once all the consequences derived above have been taken into account, one obtains the following residual the entropy inequality 
\begin{equation}
\left(\f{\p s}{\p \Phi_{q}}\right)_{,x} \Psi - \f{\p s}{\p e}q_{,x} - \f{\p s}{\p q}\Phi_{q,x} + J_{0,x}+\f{\p s}{\p q}r_q + \f{\p s}{\p \Phi_q}r_p   \geq 0 \label{resineq}
\end{equation}

A general solution of \eqref{resineq} can be achieved under additional assumptions on the entropy density and the entropy flux. Then, we assume that the entropy is a quadratic function of its nonequilibrium variables $q$ and $\Phi_q$ with constant positive coefficients $m$ and $M$, namely
\begin{equation}
s(e,q,\Phi_q)= \hat s(e) - m\f{q^2}{2} - M \f{\Phi^2_q}{2}  \label{c1}
\end{equation}
and that the following compatibility condition
\begin{equation}
\f{\p J_0}{\p \Phi_{q}} = \f{\p s}{\p q}  \label{c2}
\end{equation}
is true. We are aware that the hypotheses above are not the most general ones. However, we observe that the form 
\eqref{c1} of the entropy density ensures that the principle of maximum entropy at the equilibrium~\cite{CimJouRugVan} is fulfilled. Moreover, we will show at the end of this section, that these hypotheses  lead to a very general system of equations.

Owing to the assumptions \eqref{c1} and \eqref{c2} we get from \eqref{130a}-\eqref{130c}
\begin{equation} \label{cbis}
J_0 = -m q\Phi_q + J_1(e,q)
\end{equation}
As final condition we impose that the  entropy flux should reduce to its classical form, namely $J= \f{\p \hat s}{\p e}q $, at the equilibrium, i.e. when $\Phi_q$ and $\Psi$ do not play any role. As a consequence, we choose
\begin{equation}
J_1= \f{\p \hat s}{\p e}q \label{c3}
\end{equation}
and hence the entropy flux can be written as
\begin{equation}
J = \f{\p s}{\p \Phi_{q}} \Psi - mq\Phi_q + \f{\p \hat s}{\p e}q   \label{effin}
\end{equation}

Owing to Eqs.  \eqref{c1} and \eqref{effin}, the entropy inequality \eqref{resineq} can be simplified, and one obtains
\begin{equation}
- M \Phi_{q,x} \Psi + q\left(\left(\f{\p s}{\p e}\right)_{,x} - m r_q\right) + \Phi_q\left(\left(\f{\p s}{\p q}\right)_{,x} - M r_p\right) \geq 0 \label{resineq1}
\end{equation}

We may observe that in this quadratic expression there is a constitutive function in every term. Therefore, one may solve the inequality by linearization~\cite{Van1}, obtaining so
\begin{eqnarray}\label{v1}
\Psi & = &  -l_1 M \Phi_{q,x} \\ 
q & = & l_2\left(\left(\f{1}{T}\right)_{,x} - m r_q\right) \\\label{v2}
\Phi_q & = & l_3\left(\left(\f{\p s}{\p q}\right)_{,x} - M r_p\right) \label{v3}
\end{eqnarray}
where $l_1$, $l_2$ and $l_3$ are phenomenological coefficients and $T$ is the absolute temperature.
Looking over the derivation in one spatial dimension, we observe that in the three-dimensional case the three terms in (\ref{resineq1}) have different tensorial orders. Therefore the above linear relations are much complex, since they may have more terms. However, in the case of isotropic materials in one spatial dimension, 
the solution of \eqref{resineq1} takes the simplified form above.

The system of equations \eqref{v1}-\eqref{v3} can be coupled with the system \eqref{s1a}-\eqref{s1c} in order to eliminate the production terms, $r_q$ and $r_p$, and  the highest order flux $\Psi$. 

In this way, one obtains the following system of transport equations
\begin{eqnarray}\label{et}
e_{,t} + q_{,x} & = & 0 \\ \label{qt}
\tau_q q_{,t} + q + \tau_q \Phi_{q,x}  & = & l_2 \left(\f{1}{T}\right)_{,x} \\ 
\tau_\Phi \Phi_{q,t}  + \Phi_q - \tau_\Phi l_1 M \Phi_{q,xx}  & = & -l_3 m q_{,x} \label{fit}
\end{eqnarray}

where $\tau_q = m l_2$ and $\tau_\Phi = M l_3$. This system is similar to the one obtained in RET \cite{MR}.

\subsection{Special cases}\label{Special}
The main advantage of the procedure applied above is that it is universal, and does not require any assumption which is strictly related to the particular phenomenon to be described. It is  based on the following general assumptions
\begin{itemize}
\item the deviation from the equilibrium state is conveniently characterized by the heat flux $q$ and the flux of the heat flux  $\Phi_q$ (second-order tensor); 
\item the deviation from the classical form of the entropy current is conveniently characterized
by two tensorial functions, namely the flux of the heat flux $\Phi_q$ and the flux of $\Phi_q$ (i.e. the third-order tensor $\Psi$).
\end{itemize}

It is easy to verify that if $l_3=0$ then Eq. \eqref{fit} yields $\Phi_q=0$. As a consequence, Eq. \eqref{qt} reduces to the celebrated Maxwell-Cattaneo-Vernotte equation~\cite{Cat1}
\begin{equation}\label{32}
\tau q_{,t}+q=-k T_{,x}
\end{equation}
under the condition $\displaystyle\frac{l_2}{T^2}= k$, where $k$ is the thermal conductivity, and, for simplicity of notation, here and in the following we have put $\tau_q \equiv \tau$.
 
If, instead, $l_3 m \ne 0$, and $\tau_\Phi$ is negligible, then Eq. \eqref{fit} yields $\Phi_q=-l_3\,m\,q_{,x}$. Such a relation, when inserted in Eq. \eqref{qt}, leads to the onedimensional Guyer-Krumhansl equation~\cite{GuyKru1,GuyKru2} 
\begin{equation}\label{36}
\tau q_{,t}+q+ k T_{,x}= 3l^2 q_{,xx}
\end{equation}
provided $\tau l_3 = 3l^2$, with $l$ the mean free path of the phonons which, in crystals, are the heat carriers.
On the other hand, if in the previous equation, $q$ is negligible with respect to the other terms, one obtains Green-Naghdi type equation
\begin{equation}\label{37}
\tau q_{,t}+ k T_{,x}= 3l^2 q_{,xx}
\end{equation}
Finally, still under the hypothesis $\tau_\Phi$ negligible, if the linear constitutive equation $e=c_vT$ holds, where $c_v$ denotes the volumetric specific heat, the balance equation \eqref{et}
allows  write  $\Phi_q=-l_3 \,m \,q_{,x}=l_3c_v\,m\,T_{,t}$. As a consequence, ${\Phi_q}_{,x}=\tau l_3\,m\,c_vT_{,xt}$ and the GK equation can be put in  the Jeffrey form~\cite{Cim3} 
\begin{equation}\label{39}
\tau q_{,t}+q+ k T_{,x}= -\tau l_3mc_vT_{,xt}
\end{equation}
It is interesting to note that in all the situations analyzed above, a finite speed of propagation arises only in the MCV case. This is also the sole case in which $l_3=0$, which leads to $\Phi_q=0$. Then, Eqs. \eqref{c1} and  \eqref{effin}
reduce to the classical local expressions of EIT
\begin{equation}
s(e,q)= \hat s(e) - m\f{q^2}{2} \label{c1bis}
\end{equation}
\begin{equation}
J = \f{\p \hat s}{\p e}q   \label{effinbis}
\end{equation}

In the other (parabolic) cases we have $\tau_\Phi$ negligible, so that $s$ is still given by \eqref{c1bis} but $J$ takes the form
\begin{equation}
J = - mq\Phi_q + \f{\p \hat s}{\p e}q   \label{effinter}
\end{equation}

\section{Generalized heat-transport equation II: Hyperbolic case}\label{Equation}}

\subsection{Differential consequences of balance laws}\label{Cons}

In continuum thermodynamics the evolution of the unknown quantities is ruled by balance laws of the type
\begin{equation}
u_{,t}+{\Phi}_{u,x}= r_u \label{Cons1a}
\end{equation} 
\begin{equation} 
w_{,t}+{\Phi}_{w,x}= r_w \label{Cons1b}
\end{equation} 
where $u$ and $w$ are generic scalar state functions, $\Phi_{u}$ and  $\Phi_{w}$ are their fluxes, $r_u$ and  $r_w$ their productions, respectively. Beside Eqs. \eqref{Cons1a}  and  \eqref{Cons1b}, it is possible to consider their differential consequences, namely the higher-order differential equations obtained by differentiating the previous system with respect to space and time.

A classical example is represented by the Maxwell-Cattaneo-Vernotte system~\cite{Cat1}, namely
\begin{eqnarray}
&&e_{,t}+ q_{,x}=0 \label{Cons2a}\\
&&\tau q_{,t}+q=-k T_{,x} \label{Cons2b}
\end{eqnarray}
Under the linear assumption $e=c_vT$, where $c_v$ denotes the volumetric heat capacity, it is possible to obtain the equation ruling the evolution of the temperature, by deriving Eq.  \eqref{Cons2a} with respect to time and Eq.  \eqref{Cons2b} with respect to space. In this way one obtains
\begin{eqnarray}
&&c_vT_{,tt}+ q_{,xt}=0 \label{Cons3a}\\
&&\tau q_{,tx}+q_{,x}=-k T_{,xx} \label{Cons3b}
\end{eqnarray}
By coupling the previous equations with the balance of energy  \eqref{Cons2a}, it is possible to eliminate the heat flux and its derivatives, getting so the classical telegraph equation~\cite{Cat1}
\begin{equation}
\tau T_{,tt}+T_{,t}-\frac{k}{c_v}T_{,xx} =0 \label{Cons4}
\end{equation}

We observe that the system  \eqref{Cons2a}-\eqref{Cons2b} requires initial conditions for $T$ and $q$, while the temperature equation  \eqref{Cons4} needs initial conditions for $T$ and its time derivative. Such a second condition is, in general, assigned on physical grounds,  and the obtained solution of  \eqref{Cons4} corresponds to the solution of system  \eqref{Cons2a}-\eqref{Cons2b} which satisfies such an additional  initial condition.
Such a methodology can be applied also to derive the temperature equation  for more complex models as, for instance, the model developed in~\cite{VanFul} (see Eq. (30) therein). Here we aim to discuss for a while the possibility of substituting to the system \eqref{Cons1a}-\eqref{Cons1b} a new system obtained by substituting to some of its equations their differential consequences. For the sake of illustration, let's suppose we are applying the Cattaneo procedure illustrated above by deriving Eq.  \eqref{Cons1a}
with respect to time and Eq.  \eqref{Cons1b} with respect to space. We get so
\begin{equation}
u_{,tt}+{\Phi}_{u,xt}= r_{u,t}\label{Cons1a2}
\end{equation} 
\begin{equation} 
w_{,tx}+{\Phi}_{w,xx}= r_{w,x} \label{Cons1b2}
\end{equation} 

Preliminarily, it is worth observing that if one proves that the system \eqref{Cons1a2}-\eqref{Cons1b2} is compatible with second law of thermodynamics, then its solutions represent real processes. However, these processes could not be described by the system \eqref{Cons1a}-\eqref{Cons1b}. In fact, a solution of  \eqref{Cons1a}-\eqref{Cons1b} is a solution of  \eqref{Cons1a2}-\eqref{Cons1b2} too but, a solution of  
\eqref{Cons1a2}-\eqref{Cons1b2} does not necessarily satisfies the system  \eqref{Cons1a}-\eqref{Cons1b}. Thus, the system \eqref{Cons1a2}-\eqref{Cons1b2} admits a more general set of solutions, which contains the solutions of \eqref{Cons1a}-\eqref{Cons1b} as a subset.  However,  let's consider the Cauchy problem for both systems and let's suppose that both of them have the necessary regularity for the validity of the Cauchy-Kovaleskaya theorem. Then, it is possible to determine suitable initial conditions ensuring that a solution of system \eqref{Cons1a2}-\eqref{Cons1b2} is a solution of the system \eqref{Cons1a}-\eqref{Cons1b} too. In fact, let's suppose that
\begin{equation}
u(x,0)=u_0(x),\,\,\,w(x,0)= w_0(x) \label{Cons5}
\end{equation} 
are given initial conditions for \eqref{Cons1a}-\eqref{Cons1b}, to which corresponds a unique solution $\Big(u(x,t),\,w(x,t)\Big)$.  Then, by time derivation of $u(x,t)$ we can determine a second initial condition for Eq. \eqref{Cons1a2} by evaluating such a derivative at the time $t=0$. Moreover, being Eq. \eqref{Cons1b2} of the first order in time, no any additional initial condition for it is needed. Hence, let's consider for the system \eqref{Cons1a2}-\eqref{Cons1b2} the Cauchy problem corresponding to the initial conditions
\begin{equation}
u(x,0)=u_0(x),\,\,\, u_{,t}(x,0)= v_0(x),\,\,\,w(x,0)= w_0(x) \label{Cons6}
\end{equation} 
where $v_0(x)$ is the value assumed by $u_{,t}(x,t)$ at $t=0$. Then, the solution $\Big(u(x,t),\,w(x,t)\Big)$ of \eqref{Cons1a}-\eqref{Cons1b} satisfies the Cauchy problem for the system  \eqref{Cons1a2}-\eqref{Cons1b2} corresponding to the initial conditions  \eqref{Cons6}. Hence, by the Cauchy-Kovaleskaya theorem, it is the unique solution corresponding to the given initial conditions. This proves that any solution of a given Cauchy problem for the system \eqref{Cons1a2}-\eqref{Cons1b2} which satisfies initial conditions of the type \eqref{Cons6},  is also a solution of the system \eqref{Cons1a}-\eqref{Cons1b} satisfying initial conditions  of the type \eqref{Cons5}.
Analogously  to what said for the temperature equation, the additional initial condition $v_0(x)$ can be assigned on physical grounds, and it yields the solution of the system \eqref{Cons1a}-\eqref{Cons1b} which satisfies such an additional condition.

In order to illustrate the previous considerations by a concrete example, let's consider the following Cauchy problem
\begin{equation}
y_{,tt}= 1\label{Cons9}
\end{equation} 
\begin{equation}
y(0)=y_0,\,\,\,y_{,t}(0) = 0\label{Cons10}
\end{equation}
Such a problem admits the solution
\begin{equation}
y(t) =  \frac{t^2}{2}+ y_0 \label{Cons8}
\end{equation} 
which is also solution of the lower-order Cauchy problem
\begin{equation}
y_{,t}= t, \,\,\, y(0)=y_0 \label{Cons7}
\end{equation} 
Then, Eq.  \eqref{Cons9} can be considered as a differential consequence of the differential equation in \eqref{Cons7}, and the second of the initial conditions  \eqref{Cons10} can be determined by taking the time derivative of solution of  \eqref{Cons7}  evaluated at $t=0$.  In this way we proved that the differential equation \eqref{Cons9} with initial conditions \eqref{Cons8} 
admits the same solution of the initial value problem \eqref{Cons7}. 

The considerations above will be applied in the next Subsection in order to build up a generalized model of heat conduction.

\subsection{Generalized heat equation}\label{Gen}
In Sec. \ref{Equation1} we derived the system \eqref{et}-\eqref{fit} on a very general level and showed that it is capable to reproduce several heat-transport equations as special cases. 

An important observation is that the system  \eqref{et}-\eqref{fit} is parabolic. Such a property results evident from partial differential equation  \eqref{fit}, which is of the first order in time and second order in space. On the other hand, it is evident that such a property is directly amenable to the nonlocal constitutive equation \eqref{v1}. In fact,  a direct inspection of the system \eqref{et}-\eqref{fit} leads to the conclusion that it turns out to be hyperbolic if, and only if, one of the following two conditions occurs:
\begin{enumerate}
\item $\Phi_q=0$;
\item the constitutive equation of $\Psi$ is local.
\end{enumerate}
As we have seen in subsection \ref{Special}, the condition 1. leads to the MCV situation. Condition 2. instead, for isotropic three-dimensio\-nal  systems is forbidden by the Curie principle, since $\Psi$ and 
$\Phi_q$ have different tensorial order. 
This leads to the conclusion that, if the state space is nonlocal, in the isotropic case the Grad's 13-moments system cannot be hyperbolic. 
The observations above, and the direct compatibility with momentum series expansion of kinetic theory,  led M\"uller and Ruggeri, the early founders of RET~\cite{CimJouRugVan,DreStr,MR,Rug12a}, to the conclusion that nonlocal state spaces are not admissible, since they lead to infinite speeds of propagation of thermomechanical disturbances. Indeed, at page 2 of their book, these authors declare "Nonlocality and history have no room in extended thermodynamics"~\cite{MR}.

One of the aims of this paper is to show that such a conclusion is no longer true if we consider a different system of equations, obtained by manipulation of system 
\eqref{s1a}-\eqref{s1c}.
 
 To show that, we start by taking the derivatives of Eq. \eqref{s1b} with respect to $t$ and  of Eqs.\eqref{s1a} and  \eqref{s1c} with respect to  $x$, getting so 
\begin{eqnarray}
 &&q_{,tt}= r_{q,t}-  \Phi_{q,xt} \\ \label{3}
&& q_{,xx}=-e_{,tx}\\\label{4}
 &&\Phi_{q,tx}= r_{p,\,x}-\Psi_{,xx} \label{5}
\end{eqnarray}
Therefore, one has
\begin{equation}
q_{,tt}=  r_{q,t} - r_{p,x} +\Psi_{,xx}\label{6}
\end{equation}

Eq. \eqref{6} represents just a differential consequence of system \eqref{s1a}-\eqref{s1c}. Its right-hand side can be calculated explicitly once the constitutive equations for $\Psi$, $r_q$ and $r_p$ are known.

In order to do that, preliminarily, let's take the formal time derivative of $r_q$ and the formal space derivative of $r_p$ and $\Psi$. We get so

\begin{equation}\label{7}
 r_{q, t}=\frac{\partial{r_q}}{\partial e} e_{,t}+\frac{\partial{r_q}}{\partial e_{,x}} e_{,xt}+\frac{\partial{r_q}}{\partial q} q_{,t}+ \frac{\partial{r_q}}{\partial  {\Phi}_{q}}  {\Phi}_{{q}_{,t}}+
\frac{\partial{r_q}}{\partial q_{,xx}} q_{,xx t}
\end{equation}
\begin{equation}\label{8}
 r_{p, x}=\frac{\partial{r_p}}{\partial e} e_{,x}+\frac{\partial{r_q}}{\partial e_{,x}}e_{,xx}+\frac{\partial{r_p}}{\partial q} q_{,x}+ \frac{\partial{r_p}}{\partial {\Phi}_{q}} {\Phi}_{{q}_{,x}}+
\frac{\partial{r_p}}{\partial q_{,xx}} q_{,xx x}
\end{equation}
\begin{equation}\label{9}
\Psi_{,x}=\frac{\partial{\Psi}}{\partial e} e_{,x}+\frac{\partial{\Psi}}{\partial e_{,x}} e_{,xx}
+\frac{\partial{\Psi}}{\partial q} q_{,x}+ \frac{\partial{\Psi}}{\partial {\Phi}_{q}}  {\Phi}_{{q}_{,x}}+
\frac{\partial{\Psi}}{\partial q_{,xx}} q_{,xxx}
\end{equation}
Now we calculate $\Psi_{,xx}$ in the semilinear approximation,  i.e.  by neglecting  the terms which contain the second partial  derivatives of $\Psi$.  Thus, we get
\begin{equation}\label{10}
\Psi_{,xx}=\frac{\partial{\Psi}}{\partial e} e_{,xx}+\frac{\partial{\Psi}}{\partial e_{,x}} e_{,xxx}+\frac{\partial{\Psi}}{\partial q} q_{,xx}+ \frac{\partial{\Psi}}{\partial {\Phi}_{q}} {\Phi}_{{q}_{,xx}}+
\frac{\partial{\Psi}}{\partial q_{,xx}} q_{,xxxx }
\end{equation}
Substitution Eqs.~\eqref{7}-\eqref{10} in  Eq.~\eqref{6} yields
\begin{eqnarray}\label{11}
&& q_{,tt}+\frac{q_{,t}}{\tau}=-\frac{q}{\tau^2}+\frac{\partial{r_q}}{\partial e} e_{,t}+\frac{\partial{r_q}}{\partial e_{,x}} e_{,x\,t}+ \frac{\partial{r_q}}{\partial  {\Phi}_{q}}  {\Phi}_{{q}_{,t}}+ 
 \frac{\partial{r_q}}{\partial q_{,xx}} q_{,x x t}\\\nonumber
&&-\frac{\partial{r_p}}{\partial e} e_{,x}-
 \frac{\partial{r_p}}{\partial e_{,x}} e_{,x x}-
\frac{\partial{r_p}}{\partial {\Phi}_{q}} {\Phi}_{{q}_{,x}}
- \frac{\partial{r_p}}{\partial q_{,xx}} q_{,xx x} +
\frac{\partial{\Psi}}{\partial e} e_{,xx}+\frac{\partial{\Psi}}{\partial q} q_{,xx}\\\nonumber
&&+\frac{\partial{\Psi}}{\partial e_{,x}} e_{,xxx}+
\frac{\partial{\Psi}}{\partial {\Phi}_{q}} {\Phi}_{{q}_{,xx}}+
\frac{\partial{\Psi}}{\partial q_{,xx}} q_{,xxxx }
\end{eqnarray}
wherein we have put
 \begin{equation}\label{12}
\frac{\partial r_p}{\partial q}=\frac{ q}{ q_{,x}\tau^2},\qquad  \frac{\partial r_q}{\partial q}=-\frac 1 {\tau}
\end{equation}
with $\tau$ a relaxation time. The relations above are legitimate within the framework of EIT, where the production terms are assigned by suitable constitutive equations~\cite{CimJouRugVan}. They  are suggested by dimensional analysis,  and by the requirement that Eq.~\eqref{11} encompasses the most important heat-transport equations, such as Eqs.~\eqref{32} and~\eqref{36}.
Eq.~\eqref{11} is the generalized heat conduction equation we are looking for. With this simple restriction we introduced a convenient simplification of the consequent calculations. A more general treatment is straightforward. 

Let's explore now the thermodynamic admissibility of the thermal processes ruled by Eq.  \eqref{11}. Second law of thermodynamics requires that only those thermodynamic transformations leading to a non-negative entropy production are physically admissible. Locally, such a production reads  
    \begin{equation}
    \sigma^{(s)}={s}_{,t}+{J}_{,x}\geq 0\label{13}
    \end{equation}
 
By developing the derivatives and using Eq. \eqref{s1a}, we obtain:
\begin{eqnarray}\label{14}
&&\frac{\partial{s}}{\partial e} e_{,t}+\frac{\partial{s}}{\partial e_{,x}} e_{,xt}+\frac{\partial{s}}{\partial q} q_{,t}+ \frac{\partial{s}}{\partial  {\Phi}_{q}}  {\Phi}_{{q}_{,t}}+
\frac{\partial{s}}{\partial q_{,xx}} q_{,xx t}+
\frac{\partial{J}}{\partial e} e_{,x}\\\nonumber
&&+\frac{\partial{J}}{\partial e_{,x}} e_{,xx}-\frac{\partial{J}}{\partial q} e_{,t}+ \frac{\partial{J}}{\partial {\Phi}_{q}} {\Phi}_{{q}_{,x}}+
\frac{\partial{J}}{\partial q_{,xx}} q_{,xxx}\geq 0\nonumber
\end{eqnarray}
On the other hand, provided $\frac{\partial{\Psi}}{\partial e}\ne0$, Eq. \eqref{9} yields:
\begin{eqnarray}\label{15}
&& e_{,x}=\Big(\frac{\partial{\Psi}}{\partial e}\Big)^{-1} \Psi_{,x}-\Big(\frac{\partial{\Psi}}{\partial e}\Big)^{-1}
\frac{\partial{\Psi}}{\partial e_{,x}}e_{,xx}-\Big(\frac{\partial{\Psi}}{\partial e}\Big)^{-1}
\frac{\partial{\Psi}}{\partial q}q_{,x}\\\nonumber
&&-\Big(\frac{\partial{\Psi}}{\partial e}\Big)^{-1}\frac{\partial{\Psi}}{\partial {\Phi}_{q}} {\Phi}_{{q}_{,x}}
-\Big(\frac{\partial{\Psi}}{\partial e}\Big)^{-1}
\frac{\partial{\Psi}}{\partial q_{,xx}}q_{,xxx}\nonumber
\end{eqnarray}

To proceed further,  let's suppose that we have substituted to Eq. \eqref{qt} its differential consequence  \eqref{11}, getting so the system
\begin{eqnarray}\label{et3}
e_{,t} + q_{,x}  =  0 \\ 
q_{,tt}+\frac{q_{,t}}{\tau}=-\frac{q}{\tau^2}+\frac{\partial{r_q}}{\partial e} e_{,t}+\frac{\partial{r_q}}{\partial e_{,x}} e_{,x\,t}+ \frac{\partial{r_q}}{\partial  {\Phi}_{q}}  {\Phi}_{{q}_{,t}}+ 
 \frac{\partial{r_q}}{\partial q_{,xx}} q_{,x x t}\\\nonumber
-\frac{\partial{r_p}}{\partial e} e_{,x}-
 \frac{\partial{r_p}}{\partial e_{,x}} e_{,x x}-
\frac{\partial{r_p}}{\partial {\Phi}_{q}} {\Phi}_{{q}_{,x}}
- \frac{\partial{r_p}}{\partial q_{,xx}} q_{,xx x} +
\frac{\partial{\Psi}}{\partial e} e_{,xx}+\frac{\partial{\Psi}}{\partial q} q_{,xx}\\\nonumber
+\frac{\partial{\Psi}}{\partial e_{,x}} e_{,xxx}+
\frac{\partial{\Psi}}{\partial {\Phi}_{q}} {\Phi}_{{q}_{,xx}}+
\frac{\partial{\Psi}}{\partial q_{,xx}} q_{,xxxx }  \\ \label{qt3}
\tau_\Phi \Phi_{q,t}  + \Phi_q - \tau_\Phi l_1 m \Phi_{q,xx}   =  -l_3 q_{,x} \label{fit3}
\end{eqnarray}
At this point let's make the same hypothesis that in subsection \ref{Special} led to parabolic situations, namely $\tau_\Phi$ negligible, and, as consequence of \eqref{fit3},  $\Phi_q = -l_3 q_{,x}$. Thus, we are allowed to pursue our analysis  under the additional hypothesis $\Phi_q=c q_{,x}$, where c is a constant free parameter. 

It is clear that the situation considered here is less general with respect to that described by the system 
\eqref{et}-\eqref{fit}, since

\begin{itemize} 
\item a solution of the system \eqref{et}-\eqref{fit} is also solution of its differential consequence \eqref{et3}-\eqref{fit3} but, as we proved in Subsection \ref{Cons}, a solution  of that last system  is also solution of the former one only under opportune initial conditions;
\item we investigate the thermodynamic compatibility of \eqref{et3}-\eqref{fit3} under the additional hypothesis of negligible $\tau_\Phi$, while \eqref{et}-\eqref{fit} has been obtained in the most general case.
\end{itemize}
On the other hand, $\tau_\Phi$ negligible represents for us the most interesting situation, since we already proved that this hypothesis leads to a parabolic GK type heat-transport equation. Our aim here is to prove that this is not necessarily the case for the system \eqref{et3}-\eqref{fit3}, and that hyperbolic situations are possible. 

To this end, we continue our analysis by observing that $\Psi_{,x}= r_p -{\Phi}_{q,t}= r_p-c q_{,xt}$. Then, the relation \eqref{14} can be rewritten in the form
\begin{eqnarray}\label{15a}
&&-\frac{\partial{s}}{\partial e}q_{,x}+\frac{\partial{s}}{\partial e_{,x}}e_{,x t}
+\frac{\partial{s}}{\partial q_{,xx}}q_{,xxt}+\frac{\partial J}{\partial e}\Big(\frac{\partial \Psi}{\partial e}\Big)^{-1}r_p\\\nonumber
&&+\Big(\frac{\partial{s}}{\partial q_{,x}} 
-\frac{\partial J}{\partial e}\Big(\frac{\partial \Psi}{\partial e}\Big)^{-1} \Big)q_{,x\,t}
-\Big(\frac{\partial J}{\partial e}\Big(\frac{\partial \Psi}{\partial e}\Big)^{-1}\frac{\partial \Psi}{\partial e_{,x}}-\frac{\partial J}{\partial e_{,x}} \Big)e_{,xx}\\\nonumber
&&-\Big(\frac{\partial J}{\partial e}\Big(\frac{\partial \Psi}{\partial e}\Big)^{-1}\frac{\partial \Psi}{\partial q}-
\frac{\partial J}{\partial q}\Big)q_{,x}
-\Big(\frac{\partial J}{\partial e}\Big(\frac{\partial \Psi}{\partial e}\Big)^{-1}\frac{\partial \Psi}{\partial q_{,x}}
-\frac{\partial J}{\partial q_{,x}}\Big)q_{,xx}\\\nonumber
&&- \Big(\frac{\partial J}{\partial e}\Big(\frac{\partial \Psi}{\partial e}\Big)^{-1}\frac{\partial \Psi}{\partial q_{,xx}}- \frac{\partial J}{\partial q_{,xx}}\Big) q_{,xxx}+\frac{\partial s}{\partial q}(r_q-c q_{,xx})\geq 0\nonumber
\end{eqnarray}\label{16}
The inequality above must be satisfied whatever the thermodynamic process is, and this implies that the coefficients of those derivatives which do not belong to the state space must vanish, otherwise the inequality could be easily violated. Thus, the following thermodynamic restrictions ensue
\begin{eqnarray}\label{17}
&&\frac{\partial{s}}{\partial q_{,x}}=
\frac{\partial{J}}{\partial e}\Big(\frac{\partial \Psi}{\partial e}\Big)^{-1} \\\label{18}
&&\frac{\partial{J}}{\partial e_{,x}}= \frac{\partial{J}}{\partial e} \Big(\frac{\partial\Psi}{\partial e}\Big)^{-1}\frac{\partial \Psi}{\partial e_{,x}}\\\label{19}
&&\frac{\partial{J}}{\partial q_{,xx}}=
\frac{\partial{J}}{\partial e}\Big(\frac{\partial \Psi}{\partial e}\Big)^{-1}\frac{\partial\Psi}{\partial q_{,xx}}\\
&&\frac{\partial{s}}{\partial e_{,x}}=0,\qquad \frac{\partial{s}}{\partial q_{,xx}} =0\label{20}
\end{eqnarray}
Finally, once the relations above have been satisfied, the following reduced entropy inequality holds
\begin{eqnarray}\label{21}
&&-\frac{\partial{s}}{\partial e}q_{,x}-\frac{\partial J}{\partial e}\Big(\frac{\partial \Psi}{\partial e}\Big)^{-1}\frac{\partial \Psi}{\partial q} q_{,x}+\frac{\partial J}{\partial q}q_{,x} +\frac{\partial J}{\partial e}\Big(\frac{\partial \Psi}{\partial e}\Big)^{-1}r_p\\\nonumber
&&+\frac{\partial s}{\partial q}(r_q-c q_{,xx})-\Big(\frac{\partial J}{\partial e}\Big(\frac{\partial \Psi}{\partial e}\Big)^{-1}\frac{\partial \Psi}{\partial q_{,x}}-
\frac{\partial J}{\partial q_{,x}}\Big)q_{,xx}\nonumber
\geq 0
\end{eqnarray}
Equations \eqref{20} allow to write the volumetric entropy in the form
\begin{equation}\label{22}
s=s_0(e)-\frac 1 {2}s_1(e)q^2-\frac 1{2}s_2(e)q_{,x}^2
\end{equation}
where $ s_0(e)$ represents the equilibrium entropy while $s_1(e)$ and $s_2(e)$ are positive-definite functions of $e$, in such a way that the principle of maximum entropy at the equilibrium ~\cite{SELCIMJOU,CimJouRugVan,JCL} is fulfilled. We explicitly observe that the functions $s_1$ and $s_2$ may now depend on $e$ and are different from the constant coefficients $m$ and $M$ of Sec. \ref{Equation1}

The constitutive relation above generalizes the classical form obtained in  EIT~\cite{SELCIMJOU,CimJouRugVan,JCL}, namely
\begin{equation}\label{23}
s=s_0(e)-\frac 1 {2}s_1(e)q^2
\end{equation} 
since it contains  the  nonlocal term $-\frac 1{2}s_2(e)q_{,x}^2$. Eq. \eqref{22} yields the relations
\begin{equation} \label{24}
 \frac{\partial s}{\partial  q}=-s_1  q,\qquad \frac{\partial s}{\partial  q_{,x}}=-s_2  q_{,x}
\end{equation}  
which, owing to \eqref{17}, lead to
\begin{equation}\label{25}
\frac{\partial{J}}{\partial e}\Big(\frac{\partial \Psi}{\partial e}\Big)^{-1}=-s_2 q_{,x}
\end{equation}
In this way, Eqs. \eqref{18} and \eqref{19} take the form
\begin{eqnarray}\label{25a}
&&\frac{\partial J}{\partial e_{,x}}=-{s_2} q_{,x}\frac{\partial \Psi}{\partial e_{,x}},\\ 
&&\frac{\partial J}{\partial q_{,xx}}=-{s_2} q_{,x}\frac{\partial \Psi}{\partial q_{,xx}} \label{25b}
\end{eqnarray}
By integrating Eqs. \eqref{25a} and \eqref{25b}, we arrive to the following constitutive equation for the entropy flux
\begin{equation}\label{26}
J=J_0(e,q, q_{,x}) -{s_2}\Psi q _{,x}
\end{equation}
Finally, if we assume for $J_0(e,q, q_{,x})$ the classical local form postulated in RT~\cite{CimJouRugVan}, namely
\begin{equation}\label{27}
J_0(e,q, q_{,x})=\frac{\partial s_0}{\partial e} {q}=\frac{1}{T} {q} 
\end{equation}
where $T$ denotes the absolute temperature, we obtain
	\begin{equation}\label{28}
	J=\frac{q}{T}-{s_2}\Psi q_{,x} 
	\end{equation}
In this way, we proved that in the present model, a nonlocal entropy implies a nonlocal entropy flux. Such a nonlocality is represented 
by the additional term $\Psi q_{,x}$ which, in general, may depend on the complete set of state functions.

\section{Wave propagation}\label{Wave}
In this section we show that the general equation derived above is capable to describe propagation with finite speed even in presence of nonlocal constitutive equations. To this end, 
here we consider a special form of Eq. \eqref{qt3}, namely
\begin{equation}\label{43}
 q_{,tt}+\frac{q_{,t}}{\tau}+\frac{q}{\tau^2}+\frac{k}{c_v\tau^2}e_{,x} -\frac{3l^2}{\tau^2} q_{,x\,x}=0
\end{equation}Eq.  \eqref{fit3}
Eq. \eqref{43} can be obtained by adding to the GK equation \eqref{36} the further relaxation term $\tau^2q_{,tt}$. Thus, it is a simple generalization of the classical parabolic situation represented by the GK equation. We prove that, in this case, propagation with finite wave speeds is possible. To show that, we consider Eq.\eqref{s1a}  (local balance of energy) and Eq. \eqref{43}. In this way we get the system
\begin{eqnarray}
&&e_{,t}+ q_{,x}=0\label{44a}\\ 
 &&q_{,tt}+\frac{q_{,t}}{\tau}+\frac{q}{\tau^2}+\frac{k}{c_v\tau^2}e_{,x} -\frac{3l^2}{\tau^2} q_{,x\,x}=0
\label{44b}
\end{eqnarray}

We recall that, due to the hypothesis $\tau_\Phi$ negligible,  in the present case Eq.  \eqref{fit3} reduces to $\Phi_q= cq_{,x}$, and has been already taken into account.

In order to write Eqs. \eqref{44a}-\eqref{44b} as a first-order system, we put $q_{,t}=w$, and  $q_{,x}=z$.
Thus, system \eqref{44a}-\eqref{44b} takes the following form
\begin{eqnarray}
&&e_{,t}+ z=0\label{45a}\\
&&w_{,t}+\frac{k}{c_v\tau^2}e_{,x} -\frac{3l^2}{\tau^2} z_{,x}=-\frac{w}{\tau}-\frac{q}{\tau^2}\label{45b}\\
&&w_{,x}-z_{,t}=0\label{45c}
\end{eqnarray}

In continuum physics the systems of governing equations often my be also put in the first-order quasi-linear form
\be\label{46}
\boldsymbol{A}_0(\boldsymbol{u})\boldsymbol{u}_{,t}+\boldsymbol{A}_i(\boldsymbol{u})\boldsymbol{u}_{,{x}_i}=\boldsymbol{f}(\boldsymbol{u})
\ee
with the unknown N-column vector $\boldsymbol{u}(\boldsymbol{x},t)= (u_1,u_2,\dots u_N)^T$, where $\boldsymbol{A}_0$ and $\boldsymbol{A}_i$ are real $N \times N$ matrices and  $\boldsymbol{f}$ is a N-column vector too. The wave speeds and the amplitudes of the acceleration waves are given, respectively, by the eigenvalues $\lambda$ and the eigenvectors $\boldsymbol{r}$ of the following eigenvalue problem
\be\label{49}
(\boldsymbol{A}_i n_i-\lambda \boldsymbol{A}_0) \boldsymbol{r}=\boldsymbol{0}
\ee
The system \eqref{46} is said hyperbolic in the $t$-direction if $\text{det} \boldsymbol{A}_0 \neq 0$, and the problem  \eqref{49} has only real eigenvalues (characteristic speeds) and $N$ independents right  eigenvectors.
The system \eqref{45a}-\eqref{45c} can be re-arranged in the form \eqref{46}, with $\boldsymbol{u}=(e,w,z)^T$, $\boldsymbol{f}=(0, -\frac{w}{\tau}-\frac{q}{\tau^2}, 0)^T$ and 

\begin{equation}\nonumber
\boldsymbol{A}_0=\left[
\begin{array}{ccc}
1&0&0\\
0&1&0\\
0&0&-1
\end{array}\right]
\end{equation}

\begin{equation}\nonumber
\boldsymbol{A}_1=\left[
\begin{array}{ccc}
0&0&0\\
\frac{k}{c_v\tau^2}&0&-\frac{3l^2}{\tau^2}\\
0&1&0
\end{array}\right]
\end{equation}
The speeds of propagation of thermal disturbances are the solutions of the following characteristic equation
\begin{equation}\label{50}
det\left[
\begin{array}{ccc}
-\lambda&0&0\\
\frac{k}{c_v\tau^2}&-\lambda&-\frac{3l^2}{\tau^2}\\
0&1&\lambda
\end{array}\right]=0
\end{equation}
which can also be written as
\begin{equation}\label{51}
\lambda^3-\frac{3l^2}{\tau^2}\lambda=0
\end{equation}
Since  Eq. \eqref{51} has only the real solutions
\begin{equation}
\lambda=0\,\qquad\lambda=-\frac{l}{\tau}\sqrt{3}, \,\qquad \lambda=\frac{l}{\tau}\sqrt{3}
\end{equation}
we conclude that the system \eqref{45a}-\eqref{45c} is hyperbolic whenever $\frac{\partial \Psi}{\partial q}>0$, and  the speeds of propagation are determined by  the mean free path of the phonons and by the relaxation time. It is worth observing that it is very frequent in continuous thermodynamics that the hyperbolicity of the evolutionary systems is guaranteed  by suitable properties  of the material functions.

It is worth observing that in this purely nonlocal framework the MCV theory cannot be obtained. Such a theory can be recovered by the system \eqref{44a}-\eqref{44b} under the hypotheses  $l=0$ (absence of nonlocality) and that the terms in $\tau^2$ are negligible (very fast phenomena).

\section{Discussion}\label{Disc}

In the present paper we showed that, with the help of Grad's 13-moments system~\cite{Gra}, it is possible   to obtain two different generalized heat-transport equations. 

The first one, of the second order in space and first order in time, describes hyperbolic heat conduction only in the absence of nonlocal effects. 

The second one, of the fourth order in space and second order in time, is able to describe hyperbolic heat conduction also in the presence of nonlocal effects.

In the first case, we used the Grad's 13-moments system in its original form, and the specific internal energy, the heat flux, and the flux of heat flux as basic fields. We supposed that the spatial derivatives of these basic fields may enter the state space. 

In the second case, instead, on the same state space we have considered some differential consequences of Grad's 13-moments system, by taking the space and time derivatives of these equations. 

The first type of model is useful if one is interested to the evolution of the flux of heat flux, and/or is looking for a first order system of balance laws which can be put in symmetric form. In fact, in this case hyperbolicity and 
well-posedness of the Cauchy problem are guaranteed~\cite{CimJouRugVan,MR}.

The second type of model is important if one aims to get finite speeds of propagation even in the presence of nonlocal effects. We notice that the heat equation we used in Sec. \ref{Wave} is only a theoretical example, aimed to show that hyperbolicity and nonlocality are not incompatible. To our best knowledge, such a compatibility has never been proved before.

From the technical point of view, the novelty of the present approach with respect to that in ~\cite{VanFul,KovVan} is that here we obtained universality without need of internal variables. This renders the results more close to the kinetic theory, since in the system~\eqref{s1a}-\eqref{s1c} the heat flux has been connected to the first moment in the Grad's approximation of the Boltzmann equation~\cite{Gra}.

In future researches we aim to derive the previous generalized heat equations in the three-dimensional  case. 

We also observe that in Sec.~\ref{Equation1} we have considered the linear case in order to recover some classical heat equations. However, it is easy to see that the system \eqref{v1} could be nonlinear too, and the propagation of thermal waves could be studied also in this case.

As far as the constitutive equations \eqref{22} and \eqref{28} are concerned, it would be interesting to investigate if the nonlocal terms appearing therein could show their effects in some situations which are interesting in the applications.

For instance, in two-dimensional nanosystems, in some cases the temperature is an increasing function of the distance from a heat source. Such a situation, which seems to contradict the second law of thermodynamics, is indeed admissible in particular situations, and such an admissibility is due to nonlocal terms in the constitutive equation of the entropy flux~\cite{SELCIMJOU}.

\section{acknowledgements}

P.~R. acknowledges the financial support of the National Group of Mathematical Physics (GNFM-INdAM).

V.~P.  and K.~R. thank the support of the NKFIH 116197, NKFIH 124366 and NKFIH 123815 grants.

V.~A.~C. acknowledges the financial support of the University of Basilicata under grants RIL 2013 and RIL 2015,  and the National Group of Mathematical Physics (GNFM-INdAM).


\bibliographystyle{unsrt}       

\begin{thebibliography}{10}

\bibitem{Leb}
G.~Lebon.
\newblock Heat conduction at micro and nanoscales: {A} review through the prism
  of {E}xtended {I}rreversible {T}hermodynamics.
\newblock {\em J. Non-Equilib. Thermodyn.}, 39:35--59, 2014.

\bibitem{SELCIMJOU}
A.~Sellitto, V.~A. Cimmelli, and D.~Jou.
\newblock {\em Mesoscopic theories of heat transport in nanosystems}.
\newblock Springer, Berlin, 2016.

\bibitem{CimJouRugVan}
V.~A. Cimmelli, D.~Jou, T.~Ruggeri, and P.~V\'an.
\newblock Entropy {P}rinciple and {R}ecent {R}esults in {N}on-{E}quilibrium
  {T}heories.
\newblock {\em Entropy}, 16:1756--1807, 2014.

\bibitem{Tzo1}
D.~Y. Tzou.
\newblock A unified field approach for heat conduction from
  micro-to-macro-scales.
\newblock {\em J. Heat Trans. - T. ASME}, 117:8--16, 1995.

\bibitem{TZOU}
D.~Y. Tzou.
\newblock {\em Macro- to Microscale Heat Transfer: The Lagging Behaviour}.
\newblock Wiley, United Kingdom, second edition, 2014.

\bibitem{KovVan1}
R.~Kov\'acs and P.~V\'an.
\newblock Thermodynamical consistency of the {Dual Phase Lag} heat conduction
  equation.
\newblock {\em Contin. Mech. Thermodyn.}, 2017.
\newblock online first, arXiv:1709.06825.

\bibitem{LebJouCasMus}
G.~Lebon, D.~Jou, J.~Casas-V\'{a}zquez, and W.~Muschik.
\newblock Weakly nonlocal and nonlinear heat transport in rigid solids.
\newblock {\em J. Non-Equilib. Thermodyn.}, 23:176--191, 1998.

\bibitem{CimSelJou1}
V.~A. Cimmelli, A.~Sellitto, and D.~Jou.
\newblock Nonlocal effects and second sound in a nonequilibrium steady state.
\newblock {\em Phys. Rev. B}, 79:014303 (13 pages), 2009.

\bibitem{CimSelJou2}
V.~A. Cimmelli, A.~Sellitto, and D.~Jou.
\newblock Nonequilibrium temperatures, heat waves, and nonlinear heat transport
  equations.
\newblock {\em Phys. Rev. B}, 81:054301 (9 pages), 2010.

\bibitem{CimSelJou3}
V.~A. Cimmelli, A.~Sellitto, and D.~Jou.
\newblock Nonlinear evolution and stability of the heat flow in nanosystems:
  {B}eyond linear phonon hydrodynamics.
\newblock {\em Phys. Rev. B}, 82:184302 (9 pages), 2010.

\bibitem{VanFul}
P.~V\'an and T.~F\"{u}l\"{o}p.
\newblock Universality in heat conduction theory: weakly nonlocal
  thermodynamics.
\newblock {\em Ann. Phys. (Berlin)}, 524:470--478, 2012.

\bibitem{KovVan}
R.~Kov\'acs and P.~V\'an.
\newblock Generalized heat conduction in heat pulse experiments.
\newblock {\em Int. J. Heat Mass Transfer}, 83:613--620, 2015.

\bibitem{Cim3}
V.~A. Cimmelli.
\newblock Different thermodynamic theories and different heat conduction laws.
\newblock {\em J. Non-Equilib. Thermodyn.}, 34:299--333, 2009.

\bibitem{DreStr}
W.~Dreyer and H.~Struchtrup.
\newblock Heat pulse experiments revisited.
\newblock {\em Cont. Mech. Thermodyn.}, 5:3--50, 1993.

\bibitem{BerVan17b}
A.~Berezovski and V\'an P.
\newblock {\em Internal Variables in Thermoelasticity}.
\newblock Springer, 2017.

\bibitem{Ver}
J.~Verh\'as.
\newblock On the entropy current.
\newblock {\em J. Non-Equilib. Thermodyn.}, 8:201--206, 1983.

\bibitem{Nyi}
B.~Ny\'iri.
\newblock On the entropy current.
\newblock {\em J. Non-Equilib. Thermodyn.}, 16:179--186, 1991.

\bibitem{Cat1}
C.~Cattaneo.
\newblock Sulla conduzione del calore.
\newblock {\em Atti Sem. Mat. Fis. Univ. Modena}, 3:83--101, 1948.

\bibitem{GuyKru1}
R.~A. Guyer and J.~A. Krumhansl.
\newblock Solution of the linearized phonon {B}oltzmann equation.
\newblock {\em Phys. Rev.}, 148:766--778, 1966.

\bibitem{GuyKru2}
R.~A. Guyer and J.~A. Krumhansl.
\newblock Thermal conductivity, second sound and phonon hydrodynamic phenomena
  in nonmetallic crystals.
\newblock {\em Phys. Rev.}, 148:778--788, 1966.

\bibitem{GreNag}
A.~E. Green and P.~M. Naghdi.
\newblock A re-examination of the basic postulates of thermomechanics.
\newblock {\em Proc. R. Soc. Lond. A}, 432:171--194, 1991.

\bibitem{Gra}
H.~Grad.
\newblock On the kinetic theory of rarefied gases.
\newblock {\em Comm. Pure Appl. Math.}, 2:331--407, 1949.

\bibitem{Liu}
I-Shih Liu.
\newblock Method of {L}agrange multipliers for exploitation of the entropy
  principle.
\newblock {\em Arch. Rational Mech. Anal.}, 46:131--148, 1972.

\bibitem{Ons1}
L.~Onsager.
\newblock Reciprocal relations in irreversible processes {I}.
\newblock {\em Phys. Rev.}, 37:405--426, 1931.

\bibitem{Ons2}
L.~Onsager.
\newblock Reciprocal relations in irreversible processes {II}.
\newblock {\em Phys. Rev.}, 38:2265--2279, 1931.

\bibitem{Van1}
P.~V\'an.
\newblock Weakly nonlocal irreversible thermodynamics.
\newblock {\em Ann. Phys.}, 12:146--173, 2003.

\bibitem{CimVan}
V.~A. Cimmelli and P.~V\'an.
\newblock The effects of nonlocality on the evolution of higher order fluxes in
  nonequilibrium thermodynamics.
\newblock {\em J. Math. Phys.}, 46:112901 (15 pages), 2005.

\bibitem{MR}
I.~M\"{u}ller and T.~Ruggeri.
\newblock {\em Rational Extended Thermodynamics}.
\newblock Springer, New York, second edition, 1998.

\bibitem{ColNol}
B.~D. Coleman and W.~Noll.
\newblock The thermodynamics of elastic materials with heat conduction and
  viscosity.
\newblock {\em Arch. Ration. Mech. Anal.}, 13:167--178, 1963.

\bibitem{CimSelTri1}
V.~A. Cimmelli, A.~Sellitto, and V.~Triani.
\newblock A generalized {C}oleman-{N}oll procedure for the exploitation of the
  entropy principle.
\newblock {\em Proc. R. Soc. A}, 466:911--925, 2010.

\bibitem{JCL}
D.~Jou, J.~Casas-V\'{a}zquez, and G.~Lebon.
\newblock {\em Extended Irreversible Thermodynamics}.
\newblock Springer, Berlin, fourth revised edition, 2010.

\bibitem{Rug12a}
T.~Ruggeri.
\newblock Can constitutive equations be represented by non-local equations?
\newblock {\em Quarterly of Applied Mathematics}, LXX(3):597--611, 2012.

\end{thebibliography}

\end{document}